\documentclass[12pt]{article}
\usepackage{graphicx,amsmath,amssymb,hyperref}

\newcommand\pubnumber{Cavendish-HEP-2011-19, DAMTP-2011-70, TTK-11-42}
\newcommand\pubdate{\today}

\def\cavendish{Cavendish Laboratory, University of Cambridge, CB3 0HE,
  UK}

\def\damtp{DAMTP, University of Cambridge, CB3 0WA, UK}

\def\aachen{Institute for Theoretical Particle Physics and Cosmology,
  RWTH Aachen University, D-52056 Aachen, Germany}

\def\support{\,\footnote{Work supported by the Helmholtz Alliance
    ``Physics at the Terascale'', the Isaac Newton Trust and the
    STFC.}}

\def\speaker{\,\footnote{Speaker}}

\def\Title#1{\begin{center} {\Large #1 } \end{center}}
\def\Author#1{\begin{center}{ \sc #1} \end{center}}
\def\Address#1{\begin{center}{ \it #1} \end{center}}

\newcommand\pubblock{\rightline{\begin{tabular}{l} \pubnumber\\
         \pubdate  \end{tabular}}}
\newenvironment{Abstract}{\begin{quotation}  }{\end{quotation}}
\newenvironment{Presented}{\begin{quotation} \begin{center} 
             PRESENTED AT\end{center}\bigskip 
      \begin{center}\begin{large}}{\end{large}\end{center} \end{quotation}}

\renewcommand{\vec}[1]{\boldsymbol{#1}}

\newcommand{\newc}{\newcommand}

\def\height{2.0in}

\newc{\fb}{{\rm fb}}
\newc{\pb}{{\rm pb}}
\newc{\nb}{{\rm nb}}
\newc{\mb}{{\rm mb}}

\newc{\pt}{p_T}

\newc{\Wpm}{W^{\pm}}
\newc{\Zgam}{Z/\gamma^*}
\newc{\jpsi}{J/\psi}
\newc{\bbbar}{b\bar{b}}
\newc{\MET}{{\not\!\!E_T}}
\newc{\order}{\mathcal{O}}
\newc{\aw}{\alpha}
\newc{\as}{\alpha_S}
\newc{\mup}{\mu^+}
\newc{\mum}{\mu^-}
\newc{\mw}{M_{W}}
\newc{\sig}{\sigma}
\newc{\sigeff}{\sigma_{\rm eff}}
\newc{\ie}{i.e.~}
\newc{\cf}{c.f.~}
\newc{\pta}{p_{T\mup_1}}
\newc{\ptb}{p_{T\mum_1}}
\newc{\ptc}{p_{T\mup_2}}
\newc{\ptd}{p_{T\mup_2}}
\newc{\ord}{\mathcal{O}}

\newc{\herwig}{\texttt{Herwig++}~}
\newc{\herwigv}{\texttt{Herwig++ v2.4.2}~}
\newc{\madgraph}{\texttt{MADGRAPH}~}
\newc{\madgraphv}{\texttt{MADGRAPH v5.1.2.4}~}

\begin{document}
\begin{titlepage}
\pubblock

\vfill
\Title{Probing double parton scattering with leptonic final states at the LHC\support}
\vfill
\Author{J.~R.~Gaunt$^{a}$, C.~H.~Kom\speaker$^{\phantom{.}a,b}$, A.~Kulesza$^{c}$ and W.~J.~Stirling$^{a}$}
\Address{$^a$\cavendish,\\ $^b$\damtp,\\ $^c$\aachen}
\vfill
\begin{Abstract}
  We discuss the prospects of observing double parton scattering (DPS)
  processes with purely leptonic final states at the LHC.  We first
  study same--sign $\Wpm\Wpm$ pair production, which is particularly
  suited for studying momentum and valence number conservation
  effects, followed by discussions on double Drell--Yan and production
  of $\jpsi$ pairs.  The effects of initial state and intrinsic
  transverse momentum smearing on pair--wise transverse momentum
  balance characteristic to DPS are studied quantitatively.  We also
  present a new technique, based on rapidity differences, to extract
  the DPS component from a double $\jpsi$ sample recently studied at
  the LHCb.
\end{Abstract}
\vfill
\begin{Presented}
MPI$@$LHC 2010\\
Glasgow, UK, November 29 -- December 3, 2010
\end{Presented}
\vfill
\end{titlepage}
\def\thefootnote{\fnsymbol{footnote}}
\setcounter{footnote}{0}

\section{Introduction}

Due to the composite nature of hadrons, it is possible to have
multiple parton hard--scatterings, \ie events in which two or more
distinct hard parton interactions occur simultaneously, in a single
hadron--hadron collision.  For a given invariant mass, such cross
sections tend to increase with collision energy due to the rapidly
increasing parton fluxes when successively lower momentum fraction $x$
is being probed.  The high collision energies at the LHC thus provides
a valuable opportunity to observe multiple parton hard--scatterings.
In particular, many double parton scattering (DPS) processes involving
leptonic final states could become accessible for the first time.
These include double electroweak processes, for example production of
same--sign W pairs ($\Wpm\Wpm$) and double Drell--Yan (DDY)
interaction, and also pair production of $\jpsi$'s.  Compared with the
DPS processes already observed, namely final states involving 4 jets
(at the AFS collaboration at the CERN ISR \cite{Akesson:1986iv}), and
$\gamma$ + 3 jets (at the CDF \cite{Abe:1997xk} and the D0
\cite{Abazov:2009gc} collaborations at the Fermilab Tevatron),
properties of the leptonic final states can be measured much more
precisely.  These processes also involve different scales and initial
state partons, and hence provide complementary information on the
non--perturbative structure of the proton to the information derived
from other DPS reactions.  It is therefore important to study
properties and prospects for observing various DPS processes in
detail.

The general expression for the DPS cross section
$\sigma^{DPS}_{(A,B)}$ is given by
\begin{eqnarray}
\sigma^{DPS}_{(A,B)} &=& \frac{m}{2}\sum_{i,j,k,l}\int
dx_1dx_2dx_1'dx_2'd^2b \nonumber \\ &&
\times \Gamma_{ij}(x_1,x_2,b;t_1,t_2)\Gamma_{kl}(x_1',x_2',b;t_1,t_2)
\hat{\sigma}^A_{ik}(x_1,x_1')\hat{\sigma}^B_{jl}(x_2,x_2')\,,
\end{eqnarray}
where $\Gamma_{ij}(x_1,x_2,b;t_1,t_2)$ is the generalised double
parton distribution function for partons $i$, $j$ with momentum
fractions $x_1$, $x_2$ at scales $t_1\equiv \ln(Q_1^2)$, $t_2\equiv
\ln(Q_2^2)$. The two partons are separated by a transverse distance
$b$. The scales $t_1$ and $t_2$ are equal to the characteristic scales
of subprocesses $A$ and $B$ respectively.  The quantity $m$ is a
symmetry factor that equals 1 if $A=B$ and 2 otherwise.

For processes that probe small $x$ values, different partons may be
expected to scatter independently to a good approximation.  In this
limit, we have
\begin{eqnarray}\label{eq:dpdfs}
\Gamma_{ij}(x_1,x_2,b)&=&D_{ij}(x_1,x_2)F(b)\,,\\
D_{ij}(x_1,x_2)&=&D_{i}(x_1)D_{j}(x_2)\,,
\end{eqnarray}
where the scales are implicitly set to equal values ($t_1=t_2$).  The
first expression factorises $\Gamma_{ij}$ into a longitudinal double
parton distribution (dPDFs) $D_{ij}$ and a (flavour--independent)
transverse distribution $F(b)$.  In the second expression, $D_{ij}$ is
further factorised into two single parton distribution functions $D_i$
and $D_j$.  Using these assumptions, $\sigma^{DPS}_{(A,B)}$ can be
written as
\begin{equation}\label{eq:DPS}
  \sigma^{DPS}_{(A,B)} =
  \frac{m}{2}\frac{\sig_A\sig_B}{\sigeff}\,,\qquad\sigeff =\left[\int
    d^2b(F(b))^2\right]^{-1}\,.
\end{equation}

The quantity $\sigeff$ is expected to be energy and process dependent,
and is one of the DPS properties that requires a more precise
experimental measurement.\footnote{For concreteness, in the following
  numerical studies we shall use $\sigeff=14.5$ mb, the value obtained
  by the CDF experiment\cite{Abe:1997xk}.}  Basic sum rule
constraints, namely momentum and valence number conservation, may be
included via the longitudinal dPDFs $D_{ij}$ and could be probed in
particular processes.  In the following, we discuss the prospects of
obtaining this information from the $\Wpm\Wpm$, DDY and double $\jpsi$
processes at the LHC.  The discussion is based on
Refs.~\cite{Kom:2011nu,Kom:2011bd,Gaunt:2010pi}.  We refer readers to
other contributions to these proceedings for phenomenological studies
involving jets and recent theoretical and experimental developments.

\section{Same--sign $\Wpm\Wpm$ pair production}

It was first sugggested in Ref.~\cite{Kulesza:1999zh} that same--sign
$\Wpm\Wpm$ process might be a clean channel for the observation of
DPS. The irreducible background to the process, namely single parton
scattering (SPS) production of a $\Wpm\Wpm$ pair must be accompanied
by two additional partons in order to conserve electromagnetic charge.
These extra partons provide excellent handles to separate the DPS
signal from the SPS background.

It was subsequently pointed out in Refs.~\cite{Gaunt:2010pi,
  Gaunt:2009re} that in certain regions of final state phase space
this process could be particularly sensitive to momentum and valence
number constraints that must operate at some level on the two parton
PDFs. The relevant region is the one in which both $W$s are produced
in the {\it same} forward direction, since this configuration favours
extraction of two large $x$ (valence) quarks from one proton. Compared
with simple factorised models, the sum rule constraints would suppress
such configurations. The charged lepton $\eta$ asymmetry
\begin{equation}
a_{\eta_l}\equiv\frac{\sigma(\eta_{l_1}\times\eta_{l_2}<0)-\sigma(\eta_{l_1}\times\eta_{l_2}>0)}{\sigma(\eta_{l_1}
  \times\eta_{l_2}<0)+\sigma(\eta_{l_1}\times\eta_{l_2}>0)}\,,
\end{equation}
where $\eta_{l_i}$ are the lepton pseudo--rapidities, should hence be
positive for $|\eta_{l_1}|,|\eta_{l_2}|>\eta^{\rm min}_l$ for some
minimum pseudo--rapidity cut $\eta^{\rm min}_l$, and should also
increase with $\eta^{\rm min}_l$.

\begin{figure}[htb]
\centering
\includegraphics[height=\height]{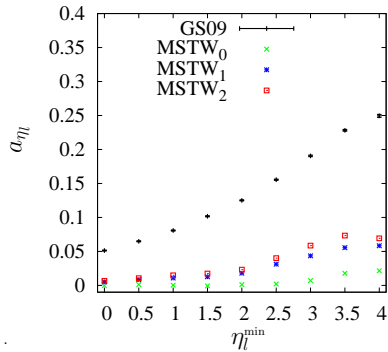}
\caption{Charged lepton pseudo--rapidity asymmetry as a function of
  $\eta^{\rm min}_l$, the minimum lepton $\eta$ cut, for
  positively--charged leptons from DPS in $pp$ collisions at
  $\sqrt{s}=14$ TeV, evaluated using different dPDF models.}
\label{fig:WWaeta}
\end{figure}

The functional dependence of $a_{\eta_l}$ on $\eta^{\rm min}_l$ is
plotted in Figure~\ref{fig:WWaeta}, assuming that proton--proton DPS
can be described in terms of dPDFs and taking various different forms
for the dPDFs.  The predictions from the GS09 dPDFs
\cite{Gaunt:2009re} in which momentum and valence number constraints
are implemented, are compared with those from simple ``MSTW$_n$''
sets, defined as $D_{ij}(x_1,x_2) =
D_i(x_1)D_j(x_2)\theta(1-x_1-x_2)(1-x_1-x_2)^n$ for $n=0,1,2$, in
which momentum constraints are implemented very crudely and valence
number constraints are not included at all.  It is known that there
are theoretical problems in describing DPS using dPDFs (see e.g.
\cite{Gaunt:2011xd}) and none of the dPDF sets used take account of
the potential contributions to DPS starting from 2 or 3
nonperturbative partons in a correct way. However, such contributions
play a subdominant role in determining the shape of
$a_{\eta_l}$($\eta^{\rm min}_l$), and the main force shaping this
distribution is in fact the inclusion of basic momentum and number sum
rule constraints, which are contained at least in an approximate way
in GS09.

\begin{table}[t]
\begin{center}
    \begin{tabular}{p{1.5in}p{0.5in}c|cc}
      Cuts&&14 TeV LHC& $\sigma_{\mu^+\mu^+}$ [fb] & $\sigma_{\mu^-\mu^-}$ [fb] \\
      \cline{1-1}\cline{3-5}
       $|\eta_l|<2.5$&&$\Wpm\Wpm$(DPS) &0.82 &0.46 \\
      $20\leq p^l_T\leq 60$ GeV&&$\Wpm \Zgam$ & 5.1 &3.6 \\
      $\MET\ge 20$ GeV&&$\Zgam \Zgam$ & 0.84 &0.67\\
      OS lepton veto&&$\bbbar\; (p^b_T\geq 20 \textrm{~GeV})$ & 0.43& 0.43\\
      \cline{1-1}\cline{3-5}
    \end{tabular}
    \caption{Selected cuts (left) on the same--sign muons, and cross
      sections [fb] (right) after cuts for signal and background
      same--sign $W$ production, including branching ratio into
      muon.\vspace{-0.5cm}}\label{tab:WW}
\end{center}
\end{table}

To see how well the DPS $\Wpm\Wpm$ signal can be extracted from
background, we perform a parton level study, relevant for the general
purpose detectors ATLAS and CMS, to look for same--sign muon pairs.
We include diboson ($\Wpm \Zgam, \Zgam\Zgam$), heavy flavour
($\bbbar$) production as well as the irreducible $\Wpm\Wpm+j$'s SPS
background discussed above.\footnote{We refer readers to the original
  paper \cite{Gaunt:2010pi} for a technical account of the
  simulation.}  The cuts and resulting cross sections are displayed in
Table~\ref{tab:WW}.  We see that a small excess can be expected, with
the background dominated by diboson production.  There are however
additional handles which can help distinguish the signal from
background.  See Ref.~\cite{Gaunt:2010pi} for more details.

\vspace{-0.3cm}

\section{DDY and double $\jpsi$ at LHCb}

One of the main characteristic in DPS is the so--called pair--wise
balancing, in which the final states from the two hard scattering
processes have zero transverse momentum at parton level.  This has
been used to help identify the presence of double parton scattering in
previous experiments.  If all four DPS final states are charged
leptons, in principle pair--wise balancing could also be observed.

In the following, we focus on events with four muon final states,
forming two opposite sign (OS) muon pairs.  We look in the low
invariant mass region, where the DPS to SPS ratio is expected to be
larger.  LHCb has excellent low $p_T$ muon acceptance, which can go
down to $\sim\,1$ GeV, and muon identification in the low mass region,
making it well suited for studying 4--muon DPS events.  In the low
mass region, the factorisation into two single 2--to--2 hard
scatterings in Eq.~\ref{eq:DPS} is likely to be a good approximation,
and so is used in the numerical analysis that follows.

We first discuss DDY.  We use \herwigv\cite{Bahr:2008pv} to generate
the DPS signal, and \madgraphv\cite{madgraph} to generate the SPS
4--muon background, which is then interfaced to \herwig for parton
showering.  We also include a Gaussian intrinsic $p_T$ smearing,
parameterised by the parameter $\sigma=2$ GeV, to provide a more
realistic simulation.  The cuts and the resulting cross sections are
displayed in Table~\ref{tab:DDY}.  We see that while at 7 TeV, the
cross section is too low for an expected integrated luminosity of
$\ord(1)$ fb$^{-1}$, it might be possible to observe DPS events at the
14 TeV LHC at high luminosity.

\begin{table}[t]
  \centering
  \begin{tabular}{p{2.1in}p{0.3in}c|r@{.}l|r@{.}l}
    Cuts &&\multicolumn{5}{c}{DDY cross sections [fb] at LHCb}\\
    \cline{1-1}\cline{3-7}
    $1.9 < \eta < 4.9$,\phantom{..} $\pt > 1$ GeV&&& \multicolumn{2}{c|}{\;\;\;\;\;\;\;\;\;DPS\;\;\;\;\;\;\;\;\;} 
    & \multicolumn{2}{c}{\;\;\;\;\;\;\;\;\;SPS\;\;\;\;\;\;\;\;\;} \\
    \cline{3-7}
    $m_{\mup\mum} > 4$ GeV&&7 TeV  & \;\;\;\;\;\;\;\;\;0&08 
    & \;\;\;\;\;\;\;\;\;0&43\\
    $9.2 < m_{\mup\mum} < 10.5$ GeV veto &&14 TeV & \;\;\;\;\;\;\;\;\;0&16 
    & \;\;\;\;\;\;\;\;\;0&68\\
    \cline{3-7}
    $ m_{4\mu} < 40$ GeV veto &&\multicolumn{5}{c}{}\\
    \cline{1-1}
  \end{tabular}
  \caption{Cuts (left) and DPS and SPS DDY cross sections [$\fb$]
    (right) for $pp$ collisions at 7 and 14 TeV.  \vspace{-0.3cm}
  }
  \label{tab:DDY}
\end{table}

\begin{figure}[htb]
\centering 
\includegraphics[height=\height]{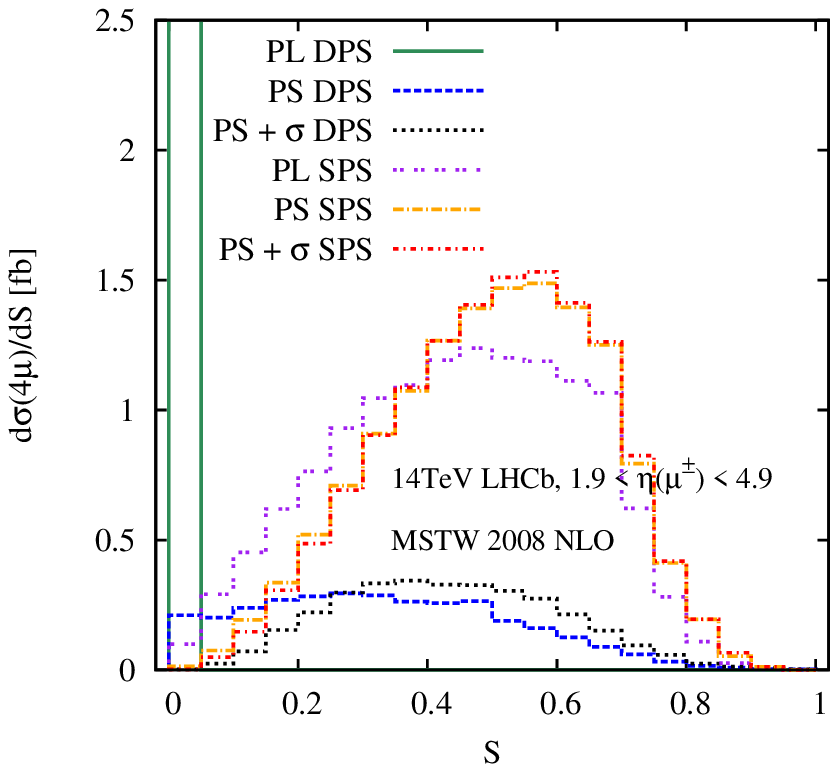}
\includegraphics[height=\height]{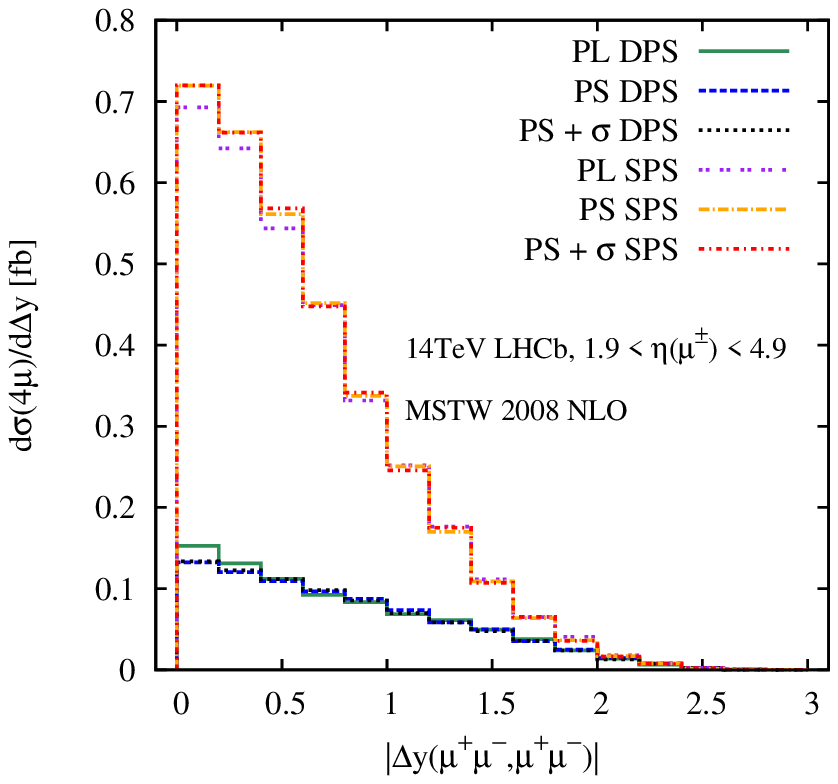}
\caption{Distributions of the pair--wise balancing variable $S$ (left)
  and the rapidity difference $|\Delta y(\mup\mum,\mup\mum)|$ (right)
  for the DPS and SPS processes including different radiation effects.
  In the figure, PL stands for ``parton level'', PS stands for
  ``parton shower'', and $\sigma$ indicates the inclusion of intrinsic
  $\pt$ smearing.}\label{fig:DDY}
\end{figure}

In principle, the DPS signal and SPS background can further be
distinguished by pair--wise balancing.  However at low invariant mass,
the presence of initial state radiation (ISR) and intrinsic $p_T$
smearing significantly affect the balancing property.  Due to the
ambiguity in grouping the 4 muons into 2 OS pairs, a pair--wise
balacing variable ($S$)
\begin{eqnarray}\label{eq:Sdef}
S&=&\frac{1}{2}\left(\frac{|\vec{\pta}+\vec{\ptb}|}{\pta+\ptb}+\frac{|\vec{\ptc}+\vec{\ptd}|}{\ptc+\ptd}\right)\,,
\end{eqnarray}
is used to group the muons into 2 OS pairs by minimising $S$.  The $S$
distributions including different radiation effects are displayed in
Figure~\ref{fig:DDY} (left plot).  Clearly, these distributions depend
sensitively on the radiation effects, with the SPS and DPS
distributions becoming more similar after these effects are included.

On the other hand, longitudinal kinematic variables are expected to be
less sensitive to the radiation effects, which primarily affects
kinematic distributions on the transverse plane.  One such variable is
the rapidity difference $|\Delta y(\mup\mum,\mup\mum)|$ between the
two OS muon pairs specified by minimising $S$, which has the
additional advantage of being invariant under longitudinal boost.  The
distributions of $|\Delta y(\mup\mum,\mup\mum)|$ are displayed in
Figure~\ref{fig:DDY} (right plot).  We see that both the DPS and SPS
distributions are much more stable against ISR and intrinsic $\pt$
smearing.  Also, the fraction of DPS events increases with $|\Delta
y(\mup\mum,\mup\mum)|$, making it an excellent variable to distinguish
DPS from SPS events.

The above observations can be applied to double $\jpsi$ production.
Compared with DDY, this process benefits from the large {\it single}
$\jpsi$ cross section, while the theoretical description of the
production is an active area of current research.  In double $\jpsi$
production there is no ambiguity in grouping the 4 muons into 2 OS
pairs, as the correct pairing should have $\mup\mum$ invariant mass
close to the physical $\jpsi$ mass.  The set of cuts and the resulting
cross section is displayed in Table~\ref{tab:DJpsi}.

\begin{table}[t]
  \centering
  \begin{tabular}{p{1.5in}p{0.5in}c|r@{.}l|r@{.}l}
    Cuts &&\multicolumn{5}{c}{Double $\jpsi$ cross sections [pb] at LHCb}\\
    \cline{1-1}\cline{3-7}
    $1.9 < \eta < 4.9$&&& \multicolumn{2}{c|}{\;\;\;\;\;\;\;\;\;DPS\;\;\;\;\;\;\;\;\;} 
    & \multicolumn{2}{c}{\;\;\;\;\;\;\;\;\;SPS\;\;\;\;\;\;\;\;\;} \\
    \cline{3-7}
     $\pt > 1$ GeV&&7 TeV  & \;\;\;\;\;\;\;\;\;3&16 
    & \;\;\;\;\;\;\;\;\;1&70\\
     $m_{\mup\mum} \simeq m_{\jpsi}$&&14 TeV & \;\;\;\;\;\;\;\;\;7&69 
    & \;\;\;\;\;\;\;\;\;2&62\\
    \cline{1-1}\cline{3-7}
  \end{tabular}
  \caption{Cuts (left) and DPS and SPS double $\jpsi$ cross sections
    [$\pb$] (right) including ${\mathcal BR}(\jpsi\to\mup\mum)$ for
    $pp$ collisions at 7 and 14 TeV. \vspace{-0.5cm}
  }
  \label{tab:DJpsi}
\end{table}

\begin{figure}[htb]
\centering 
\includegraphics[height=\height]{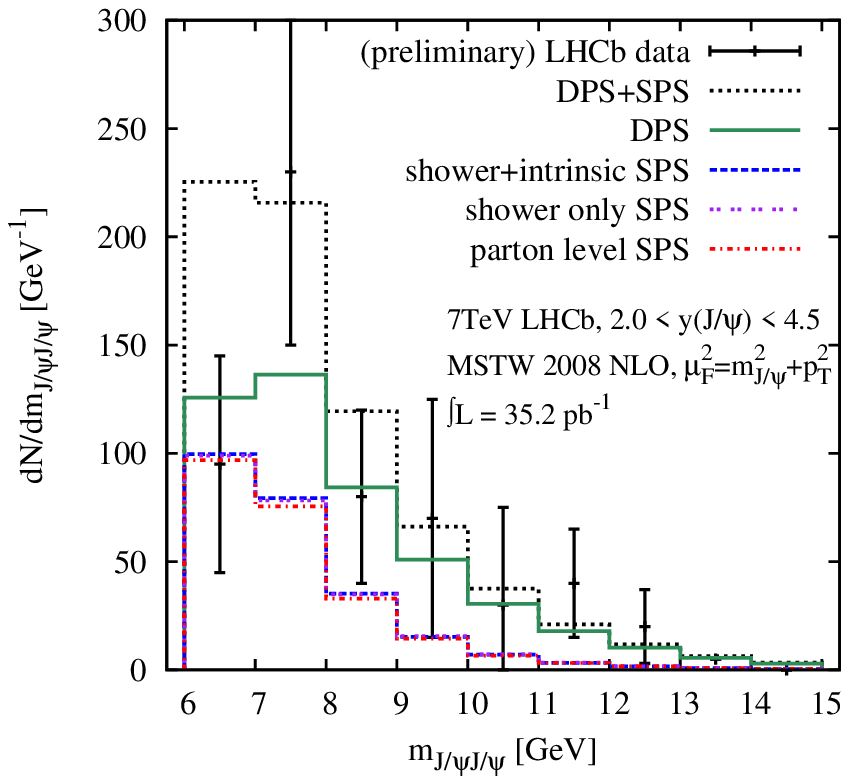}
\includegraphics[height=\height]{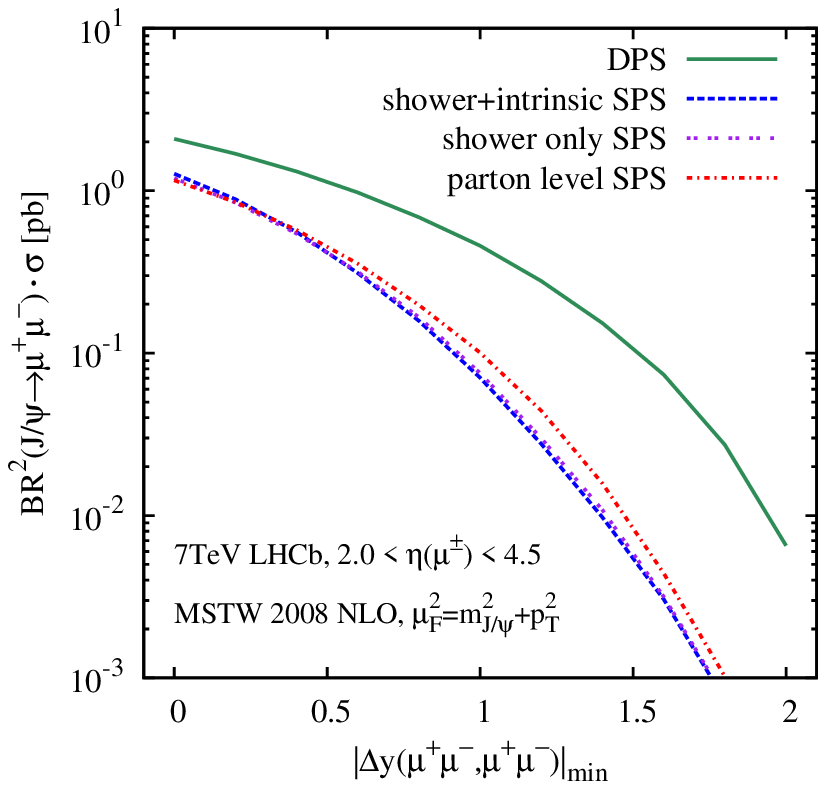}
\caption{Distributions of invariant mass $m_{\jpsi\jpsi}$ (left) and
  variation of the cross section as a function of minimum rapidity
  difference $|\Delta y(\mup\mum,\mup\mum)|$ (right) for the DPS and
  SPS processes.  Note that slightly different cuts, motivated by the
  cuts used in the recent LHCb measurement
  \protect{\cite{LHCb-CONFNote}}, have been used to obtain the above
  distributions, \cf\protect{\cite{Kom:2011nu,Kom:2011bd}}.}
\label{fig:DJpsi}
\end{figure}

In fact, recent results from LHCb \cite{LHCb-CONFNote} might already
indicate the presence of double $\jpsi$ events from DPS.  Assuming
$\sigeff=14.5$ mb, the theoretical cross sections for the SPS and DPS
double $\jpsi$ processes are similar.  However the $m_{\jpsi\jpsi}$
invariant mass distribution is different, with the DPS distribution
peaking at slightly higher values.  In Figure~\ref{fig:DJpsi} (left
plot), we show that combining contributions from DPS and SPS might
provide a better fit to data.

The large double $\jpsi$ cross section allows extraction of the DPS
events from the SPS background by imposing a cut on minimal $|\Delta
y(\mup\mum,\mup\mum)|$.  In Figure~\ref{fig:DJpsi} (right plot), we
show the variation of cross section as a function of min $|\Delta
y(\mup\mum,\mup\mum)|$.  Increasing min $|\Delta
y(\mup\mum,\mup\mum)|$ can result in much higher DPS fractions.  With
more upcoming LHC data, this could be an excellent tool in
establishing the presence of DPS double $\jpsi$ events \cite{Kom:2011bd}.

\vspace{-0.5cm}

\section{Summary}

Measurements undertaken at the LHC will be crucial for improving the
theoretical description of multiple parton scattering, in particular
double parton scattering.  We have studied the prospects of observing
DPS with leptonic final states.  Both same--sign $\Wpm\Wpm$ and
Drell--Yan pair production can be considered standard candle processes
at hadron colliders, as the leptonic final states provide clean
experimental signatures, while the theoretical predictions of the
2--to--2 subprocesses are under control.  The cross sections for these
processes are of $\ord(0.1)$ fb, and so will require high luminosities
to obtain unambiguous signals.  On the other hand, double $\jpsi$
production has a much larger cross section, and a significant DPS
component may already be present in a recent LHCb study.  With more
data in the 7 TeV run, the DPS component can be disentangled from the
SPS component using the rapidity separation between the two
reconstructed $\jpsi$'s.

\vspace{-0.3cm}



\begin{thebibliography}{99}



\bibitem{Akesson:1986iv}
  T.~Akesson {\it et al.}  [Axial Field Spectrometer Collaboration],
  Z.\ Phys.\  C {\bf 34} (1987) 163.

\bibitem{Abe:1997xk}
  F.~Abe {\it et al.}  [CDF Collaboration],
  Phys.\ Rev.\  D {\bf 56} (1997) 3811.

\bibitem{Abazov:2009gc}
  V.~M.~Abazov {\it et al.}  [D0 Collaboration],
  Phys.\ Rev.\  D {\bf 81} (2010) 052012
  [arXiv:0912.5104 [hep-ex]].

\bibitem{Kom:2011nu}
  C.~H.~Kom, A.~Kulesza, W.~J.~Stirling,
  [arXiv:1109.0309 [hep-ph]].

\bibitem{Kom:2011bd}
  C.~H.~Kom, A.~Kulesza, W.~J.~Stirling,
  Phys.\ Rev.\ Lett.\  {\bf 107} (2011) 082002
  [arXiv:1105.4186 [hep-ph]].

\bibitem{Gaunt:2010pi}
  J.~R.~Gaunt, C.~H.~Kom, A.~Kulesza, W.~J.~Stirling,
  Eur.\ Phys.\ J.\  {\bf C69}, 53-65 (2010).
  [arXiv:1003.3953 [hep-ph]].

\bibitem{Kulesza:1999zh}
  A.~Kulesza and W.~J.~Stirling,
  Phys.\ Lett.\  B {\bf 475} (2000) 168
  [arXiv:hep-ph/9912232].

\bibitem{Gaunt:2009re}
  J.~R.~Gaunt and W.~J.~Stirling,
  JHEP {\bf 1003} (2010) 005
  [arXiv:0910.4347 [hep-ph]].

\bibitem{Gaunt:2011xd}
  J.~R.~Gaunt, W.~J.~Stirling,
  JHEP {\bf 1106}, 048 (2011).
  [arXiv:1103.1888 [hep-ph]].

\bibitem{Bahr:2008pv}
  M.~Bahr, S.~Gieseke, M.~A.~Gigg, D.~Grellscheid, K.~Hamilton, O.~Latunde-Dada, S.~Platzer, P.~Richardson {\it et al.},
  Eur.\ Phys.\ J.\  {\bf C58}, 639-707 (2008).

\bibitem{madgraph}
  J.~Alwall, M.~Herquet, F.~Maltoni, O.~Mattelaer, T.~Stelzer,
  [arXiv:1106.0522 [hep-ph]];
  J.~Alwall, P.~Demin, S.~de Visscher, R.~Frederix, M.~Herquet, F.~Maltoni, T.~Plehn, D.~L.~Rainwater {\it et al.},
  JHEP {\bf 0709}, 028 (2007).
  [arXiv:0706.2334 [hep-ph]];
  F.~Maltoni, T.~Stelzer,
  JHEP {\bf 0302}, 027 (2003).
  [hep-ph/0208156].

\bibitem{LHCb-CONFNote} 
  LHCb Collaboration, 
  LHCb-CONF-2011-009.
  [arXiv:1109.0963 [hep-ex]].


\end{thebibliography}
\end{document}